\documentclass[journal,twocolumn,compsoc]{IEEEtran}
\usepackage{cite}
\usepackage{amsmath,amssymb,amsfonts}
\usepackage{algorithmic}
\usepackage{graphicx}
\usepackage{textcomp}

\usepackage[whole]{bxcjkjatype}
\usepackage[hyphens]{url}
\newcommand{\bhline}[1]{\noalign{\hrule height #1}}
\usepackage{seqsplit}
\usepackage{multirow}
\usepackage{textcomp}
\usepackage{colortbl,array,xcolor}

\usepackage{titlesec}
\usepackage{caption}

\titleformat{\paragraph}[hang]
  {\normalfont\normalsize\bfseries} 
  {\theparagraph}
  {1em}
  {}
  []

\titlespacing*{\paragraph}
  {0pt}
  {3.25ex plus 1ex minus .2ex}
  {0em}

\begin{document}

\title{Binaural Unmasking in Practical Use: \\
Perceived Level of Phase-inverted Speech in Environmental Noise}

\author{
  \IEEEauthorblockN{Rina Kotani, Chiaki Miyazaki, and Shiro Suzuki \\}
  \IEEEauthorblockA{Sony Group Corporation\\
  Tokyo, Japan\\
 \{rina.kotani, chiaki.miyazaki, shiro.suzuki\}@sony.com}
}

\maketitle

\begin{abstract}
We aim to develop a technology that makes the sound from earphones and headphones easier to hear without increasing the sound pressure or eliminating ambient noise. To this end, we focus on harnessing the phenomenon of binaural unmasking through phase reversal in one ear. Specifically, we conduct experiments to evaluate the improvement of audibility caused by the phenomenon, using conditions that approximate practical scenarios. We use speech sounds by various speakers and noises that can be encountered in daily life (urban environmental sounds, cheers) to verify the effects of binaural unmasking under conditions close to practical situations. The results of experiments using the Japanese language showed that (i) speech in a noisy environment is perceived to be up to about 6 dB louder with phase reversal in one ear, and (ii) a certain effect (improvement of audibility by 5 dB or more) is obtained for all speakers and noises targeted in this study. These findings demonstrate the effectiveness of binaural unmasking attributed to interaural phase differences in practical scenarios.
\end{abstract}

\section{Introduction}
\label{sec:intro}
We aim to develop a technology that makes the sound from earphones and headphones easier to hear without increasing the sound pressure or eliminating ambient noise. Users of earphones and headphones may be at risk of hearing loss if they continue to listen to sound at high sound pressure levels \cite{who:2015}, so reducing the sound pressure is important. The straightforward method to make the sound from earphones and headphones (i.e., the target sound) more audible in a noisy environment is to eliminate the noise masking the target sound (e.g., by noise cancellation). However, there are certain noises that should not be eliminated for safety and communication reasons, such as the sound of approaching cars or station announcements. Thus, a strategy that makes the target sound more audible without eliminating noise is needed. In other words, we need a technology that enables one to ``listen to the target sound while hearing the surrounding sounds.'' Open-ear earphones that do not block out ambient sounds by not plugging the ear canal have been developed and are becoming more available
\cite{sony:linkbuds, bose:earbuds, huawei:freeclip}, but there remains the problem that the target sound becomes inaudible when the surrounding noise is too loud \cite{headphonecheck:huawei}.

Phenomena that make the target sound more audible in noisy environments include {\em binaural unmasking for detection} \cite{hirsh:1948} and {\em binaural gain in intelligibility} \cite{licklider:1948}, which are caused by interaural phase differences, especially phase reversal in one ear. It has been suggested that phase reversal in the low-frequency range (i.e., below 500 Hz ) is important for speech \cite{levitt:1967}. In addition, studies have been done to confirm the effect of binaural unmasking using speech in various languages (such as English \cite{levitt:1967}, Chinese \cite{ho:2015}, Swedish \cite{johansson:2002} and Persian \cite{ashrafi:2022}).

However, most of the speech that has been studied so far has been read by a single male \cite{levitt:1967, ho:2015, johansson:2002, ashrafi:2022}, and there is little research involving various  speakers with different formant patterns, including women. In other words, the effect on a variety of speakers is not clear. Moreover, the noises that have been studied are mostly mechanically generated ones such as broad-band Gaussian noise \cite{levitt:1967}, speech spectrum noise \cite{ho:2015}, and speech weighted noise \cite{johansson:2002}. Therefore, the effect on  noises found in everyday life (such as urban environmental sounds) is also not clear.

Although the phenomenon  of binaural unmasking attributed to interaural phase differences is well-known, the feasibility for engineering applications has yet to be sufficiently examined. In this study, we conduct the following investigations to assess the effectiveness of binaural unmasking through phase reversal in one ear, with the aim of applying this phenomenon to practical use.

\begin{itemize}
\item Investigation of the effect for a variety of speakers by using speakers with different formant patterns, including women.
\item Investigation of the effect for a variety of noises by using noises in everyday life, such as urban environmental sounds and cheers.
\item Analysis of the relationship between the effect and frequency bands by conducting phase reversal limited to specific frequency ranges, referencing the previous study \cite{levitt:1967}. 
\end{itemize}

\begin{figure}[btp]
\centering
\includegraphics[width=\linewidth]{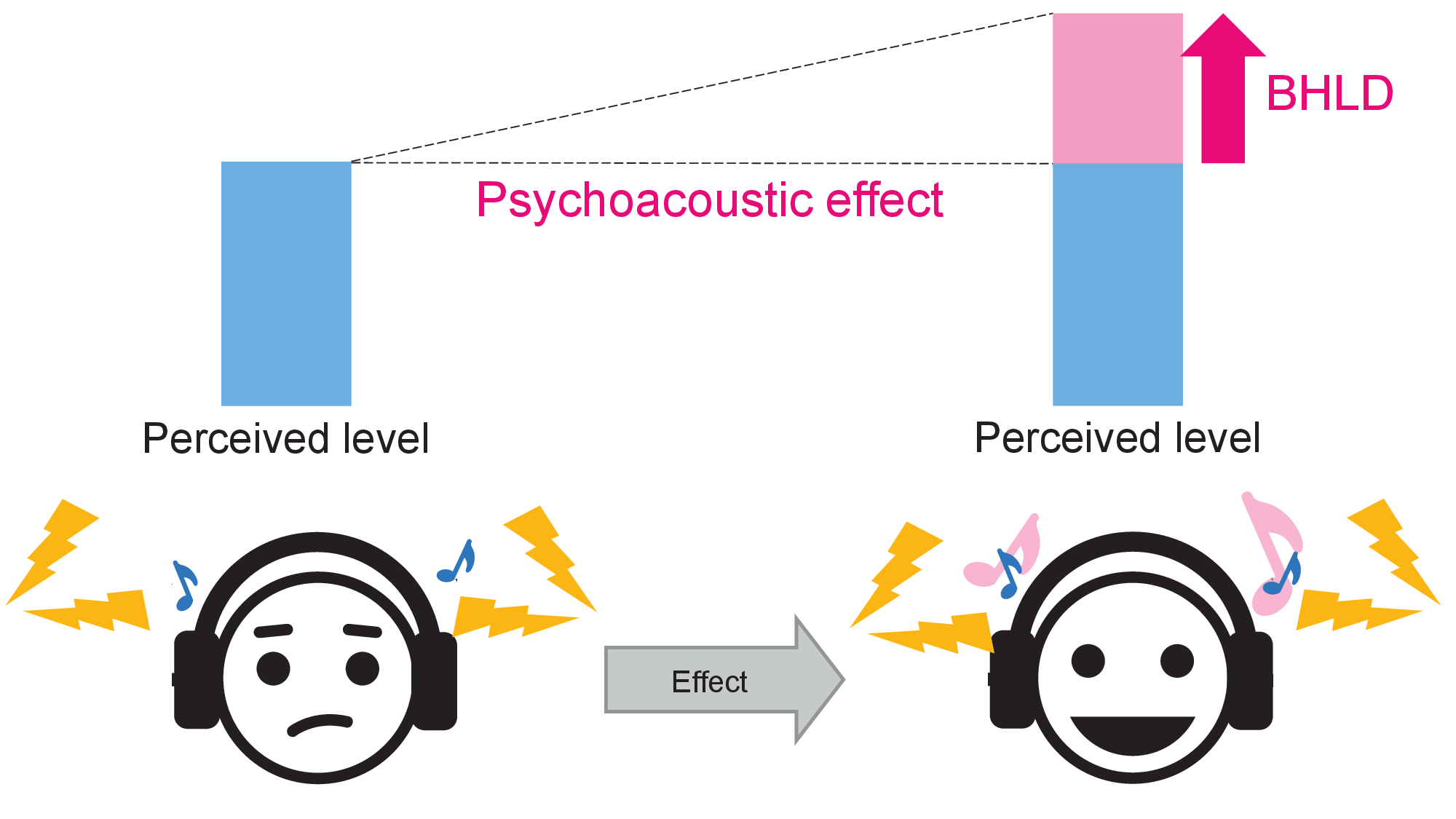}
\vspace{-5mm}
\caption{Schematic diagram of BHLD.}
\label{bhld}
\end{figure}

Previous studies \cite{levitt:1967, ho:2015, johansson:2002, ashrafi:2022} have evaluated the effect of this phenomenon using a metric called {\em Binaural Masking Level Difference} (BMLD), which refers to the improvement in detection thresholds when interaural differences are introduced in the stimuli \cite{anderson:2018}. When considering practical scenarios, however, what we want to know is not ``the barely audible volume of speech in noise'' but rather ``the improvement in hearing in a difficult-to-hear situation.'' In this paper, we introduce a straightforward method to measure this improvement. In terms of focusing on hearing (i.e., how well the target sound can be heard) rather than masking (i.e., how much it is masked), we refer to the metric used in this paper as the {\em Binaural Hearing Level Difference} (BHLD), which represents the amount of improved audibility due to perceptual effects. A conceptual diagram of BHLD is provided in Fig. \ref{bhld}.

Note that this study focuses on the improvement in perceived level, and the improvement in understanding the content being spoken (i.e., intelligibility) is not dealt with. The target language of this paper is Japanese, which is the native language of the authors.

To summarize, the novelty of this paper lies in (i) proposing the utilization of the phenomenon of binaural unmasking attributed to interaural phase differences as a new approach to improving the audibility of earphones and headphones, (ii) evaluating the effect of this phenomenon under experimental conditions that approximate real-world environments, and (iii) introducing a new evaluation metric for assessing the practical effects of audibility improvement. 

\section{Related Work}
\label{sec:relwork}
\subsection{Engineering Approaches for Better Hearing}
\subsubsection{Noise Cancellation}
The straightforward approach to making a target sound more audible in a noisy environment is to eliminate the noise masking the target sound, that is, {\em noise cancellation}. There are two types of method for noise cancellation \cite{manisha:2019}: {\em Passive noise cancellation}, which blocks ambient noise by using materials in the headphones that absorb and reflect sound \cite{shalool:2016}, and {\em Active noise cancellation}, which eliminates noise by applying ``anti-noise'' of the same amplitude but with an inverted phase \cite{kuo:2006}. Although these techniques are effective for making the target sound more audible in noisy environments, they do not meet our objective of ``listening to the target sound while hearing the surrounding sounds'' because they eliminate ambient noise.

\subsubsection{Speech Enhancement and Speech Restoration}
The technology that helps in refining the original speech signal to a higher quality is called {\em speech enhancement} \cite{taha:2018}, and it is currently utilized in devices such as hearing aids \cite{ravi:2016}. However, this technology also eliminates sounds other than speech, which does not align with our objective. There is a similar technology called {\em speech restoration} \cite{ravi:2016}, which specifically aims to reconstruct and restore the signal after degradation without altering the original speech. Yet, this technology also does not meet our objective because it eliminates sounds other than speech.

\subsection{Scientific Findings on Better Hearing}
\subsubsection{Binaural Unmasking}
{\em Binaural unmasking} refers to the phenomenon that it is substantially easier to detect a signal in noise when the interaural parameters of the signal are different from those of the noise \cite{dieudonne:2024}.

Hirsh's experiments in 1948 revealed that under intense noise, monaural listening makes it easier to detect pure tones than binaural listening \cite{hirsh:1948a}. However, further experiments by Hirsh demonstrated that the detection threshold becomes lower when a target sound presented binaurally, masked by a noise identical at the ears, is inverted in one ear compared to when presented monaurally \cite{hirsh:1948}. It was also shown that this effect is most pronounced around the 200–500-Hz range and that the greatest reduction in detection threshold occurs when a target sound is presented 180\textdegree out-of-phase interaurally \cite{hirsh:1948}.

When listening to a 500-Hz  pure tone in a noise identical at the ears, inverting the phase of the target sound in one ear can lower the detection threshold by about 12–15 dB \cite{hirsh:1948a}. However, this effect declines to about 3 dB as the frequency of the target sound increases, and stabilizes at about 3 dB for all higher frequencies, up to at least 4 kHz \cite{hirsh:1958}. The effect of binaural unmasking becomes more pronounced as the masking noise level increases, that is, as the signal-to-noise ratio (SNR) decreases, and it diminishes as the SNR increases \cite{hirsh:1948, johansson:2002, blauert:2001, culling:2006, eva:2023}.

Not only the detection of pure tones but also the performance of word articulation tests (i.e., intelligibility) improves due to phase reversal between the ears \cite{licklider:1948}. Regarding the detection of speech in noise and word articulation, experiments that limited the frequency range targeted for phase reversal showed that phase reversal in the low-frequency range (below 500 Hz) is important \cite{levitt:1967}. Furthermore, the effects of binaural unmasking due to interaural phase differences have been investigated with speech in various languages, including Chinese \cite{ho:2015}, Swedish \cite{johansson:2002}, and Persian \cite{ashrafi:2022}.

As discussed above, the phenomenon of binaural unmasking has been studied for a long time. However, as mentioned in Section \ref{sec:intro}, the effectiveness of the phenomenon for a variety of speakers and noises is still not clear, indicating that verification for practical applications is required.

\subsubsection{Physiological Findings on BMLD}
The mechanisms underlying the BMLD have been partially revealed through physiological experiments using guinea pigs.

The integration of sounds between the ears is first carried out in the {\em superior olivary complex} (SOC) located in the brainstem \cite{moore:1991}. When sounds 180\textdegree out-of-phase interaurally are presented to each ear, an {\em Interaural Time Difference} (ITD) occurs due to the phase shift between the ears. It is suggested that the SOC, responding to this ITD, processes the phase differences as cues for directional angles, which may help separate the target sound from noise by placing it at a different perceived location, thus making it more audible \cite{jiang:1997}.

When listening to a composite sound of noise and a target sound, the firing patterns of neurons in the {\em inferior colliculus} (IC), located in the midbrain, differ between when both the noise and the target sound are in phase at both ears ($\rm{N_0S_0}$) and when the noise is in phase but the target sound is 180\textdegree out of phase ($\rm{N_0S_{\pi}}$). These differences in firing pattern are said to be involved in the BMLD \cite{jiang:1997, jiang:1997b}. They are also consistent within the auditory cortex, which is further upstream from the IC \cite{gilbert:2015}.

\section{Experiment}
\label{sec:exp}
\subsection{Purpose and Overview}
\label{sec:purpose}
We conducted an experiment to determine the effects of binaural unmasking based on interaural phase differences using experimental conditions that simulate practical scenarios. The language targeted in this experiment was Japanese.

In the experiment, participants listened through headphones to an original signal (Original: $\rm{S_0}$) and a signal in anti-phasic conditions (Comparison: variations of $\rm{S_{\pi}}$), both played with an identical noise at both ears   ($\rm{N_0}$). The initial gain of the Original signal (i.e., the gain before the volume adjustment) had been determined with the intention of creating an initial state of ``audible but difficult to hear'' through the process described in Appendix \ref{sec:initgain}. The participants were asked to compare the two sounds and adjust the volume of the Original until it was perceived as having the same level as the Comparison. In most cases, the Comparison was more audible than the Original, which means the volume of the Original needed to be increased to match the Comparison. If the Comparison was less audible, the volume of the Original was decreased. By comparing the two signal levels before and after the adjustment, we measured the change in perceived level of the speech in noise resulting from the phase reversal of the speech in one ear. Details will be described later in Section \ref{sec:proc}.

This experiment involved human participants. Approval of all ethical and experimental procedures and protocols was granted by Sony Bioethics Committee (no. 25-R-0011).

\subsection{Participants}
\label{sec:part}
The experiment was conducted with 17 participants; however, one participant was excluded for the reason described later in Section \ref{sec:res}. Therefore, the experimental results were compiled based on data from 16 participants. The 16 participants were recruited from a corporate group that includes the authors' company, excluding the authors themselves. Among them, eleven were male and five were female, with two in their 20s, seven in their 30s, four in their 40s, and three in their 50s. All participants were native speakers of Japanese. Participants were limited to those who had no history of hearing issues in past hearing tests. However, note that we did not ask participants to submit the results of hearing tests; they only self-reported that they had never experienced hearing issues. No compensation or other benefits were provided to the participants, and the decision to participate was left to their free will. Each participant took part in the experiment only once.

\subsection{Procedure}
\label{sec:proc}
\begin{figure}[btp]
\centering
\includegraphics[width=\linewidth]{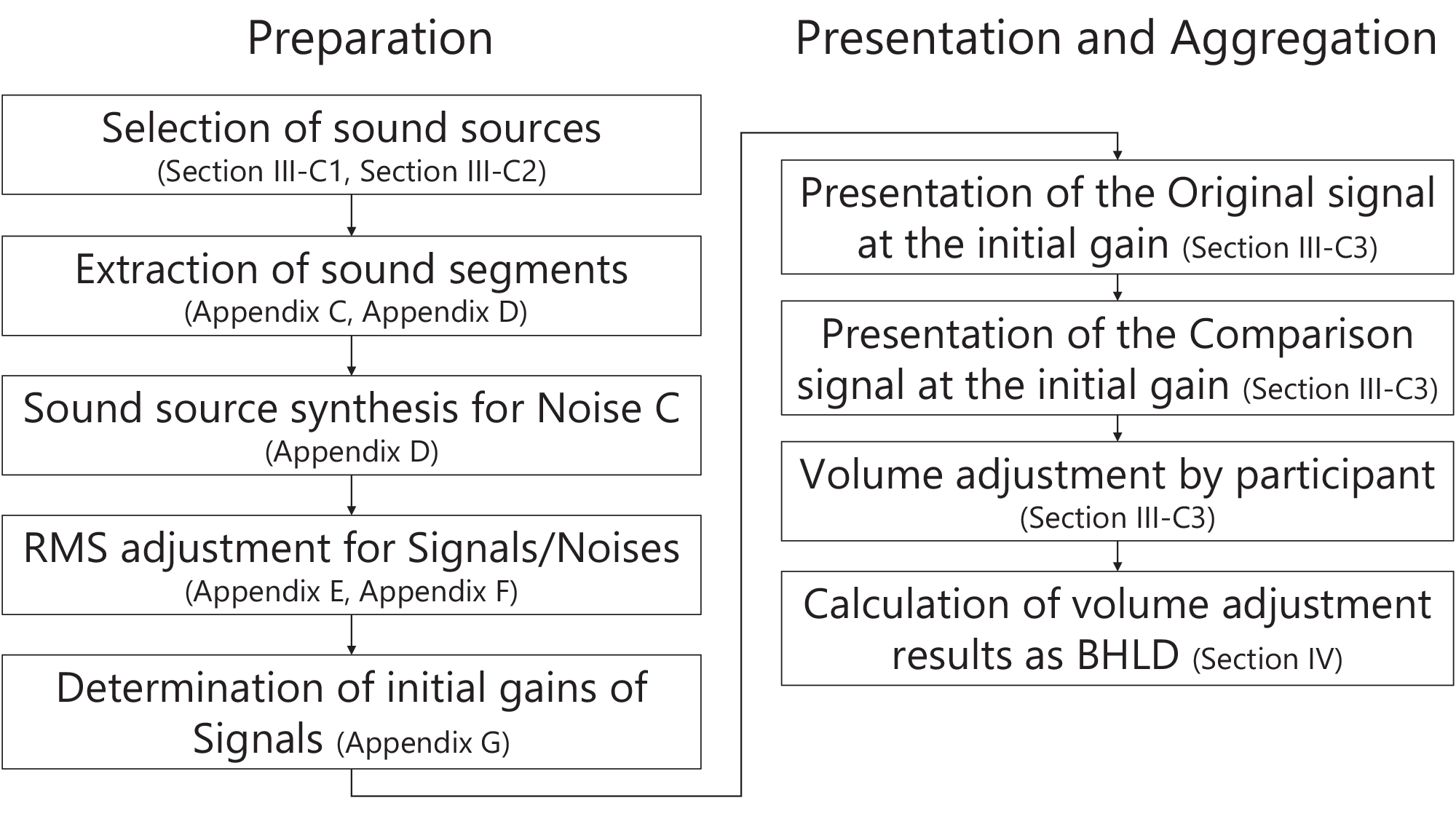}
\vspace{-6mm}
\caption{Procedure of the experiment.}
\label{procedure}
\end{figure}

Fig. \ref{procedure} shows the procedure of the experiment, which consists of two parts: ``preparation'' and ``presentation and aggregation.'' First, in the preparation part, sound sources were chosen and processed for creating Signals/Noises to be presented to the participants. In addition, initial gains of the Signals were determined. Then, the Signals/Noises were presented to the participants and they adjusted the volume of the Original signal by comparing it with the Comparison signal. Finally, the results of the volume adjustment were calculated as the BHLD .

\subsubsection{Selection of Speech Signals}
\label{sec:signal}
The content of the speech used in the experiment was a sentence in Japanese from the ITA corpus\footnote{\url{https://github.com/mmorise/ita-corpus}}, specifically \texttt{\seqsplit{RECITATION324\_180}} ``日本へ行くには船か飛行機が必要です,'' which means ``To go to Japan, you need a ship or an airplane.'' The ITA corpus is a collection of sentences selected and modified for readability and public decency from literary works whose copyrights have expired \cite{koguchi:2021}. \texttt{\seqsplit{RECITATION324\_180}} was chosen because it meets the following criteria, with the aim of selecting a sentence that would not distract the participants' concentration:

\begin{itemize}
\item Not too long: Less than 30 characters
\item In an ordinary speaking style: The polite form
\item Composed of general words: No proper nouns other than country and region names, no technical terms, no onomatopoeias
\item Easy to understand: Not a sentence explaining an extraordinary situation or one with unfamiliar premises
\end{itemize}

The speech signals utilized in the experiment were the recordings of \texttt{\seqsplit{RECITATION324\_180}} spoken by narrators or voice actors (one male, two females). Specifically, we used \texttt{male/02\_fast} and \texttt{female/02\_fast} from SpeedSpeech-JA-2022\footnote{\url{https://ast-astrec.nict.go.jp/release/speedspeech_ja_2022/}}, and the ``normal'' version from Amitaro's Voice Material Studio\footnote{\url{https://amitaro.net/}}. Hereafter, we refer to them as Signals A, B, and C, respectively. Signal A is a male voice, while Signals B and C are female voices with different formant patterns. The fact that two out of the three speakers were female was a result of using three speakers with different formant patterns. The spectrograms of Signals A, B, and C are shown in Fig. \ref{spec-signal}. It can be observed here that the voice of the second female speaker (Signal C) contains stronger high-frequency components. The duration of Signals A, B, and C is approximately 3 seconds each. The ``fast'' versions from SpeedSpeech-JA-2022 were chosen because the ``normal'' speaking rate felt slower than what is typically heard in everyday life. The details of how Signals A, B, and C were processed are described in Appendix \ref{align-speech}. Licensing information for the sound sources is listed in Table \ref{resourcelist} of the Appendix.

\subsubsection{Selection of Noises}
\label{sec:noise}
We used white noise, cheers, and urban environmental sounds, referred to as Noises A, B, and C, respectively. Noise A was generated using a random number generator with uniform distribution of MATLAB\footnote{\url{https://www.mathworks.com/products/matlab.html}}. Noise B was created by extracting a portion of the sound source found in ``歓声・拍手 (cheers and applause)'' of OtoLogic\footnote{\url{https://otologic.jp/}}. Noise C was a mix of five types of sounds from ``Environment 2'' of TK'S FREE SOUND FX\footnote{\url{https://taira-komori.jpn.org/freesounden.html}}. The details of how Noises A, B, and C were created are described in Appendices \ref{sec:noisea}, \ref{sec:noiseb},  and \ref{sec:noisec}, respectively. Licensing information for the sound sources is listed in Table \ref{resourcelist} of the Appendix.

\subsubsection{Presentation of Signals/Noises and Volume Adjustment by Participants}
\begin{figure}[btp]
\centering
\includegraphics[width=\linewidth]{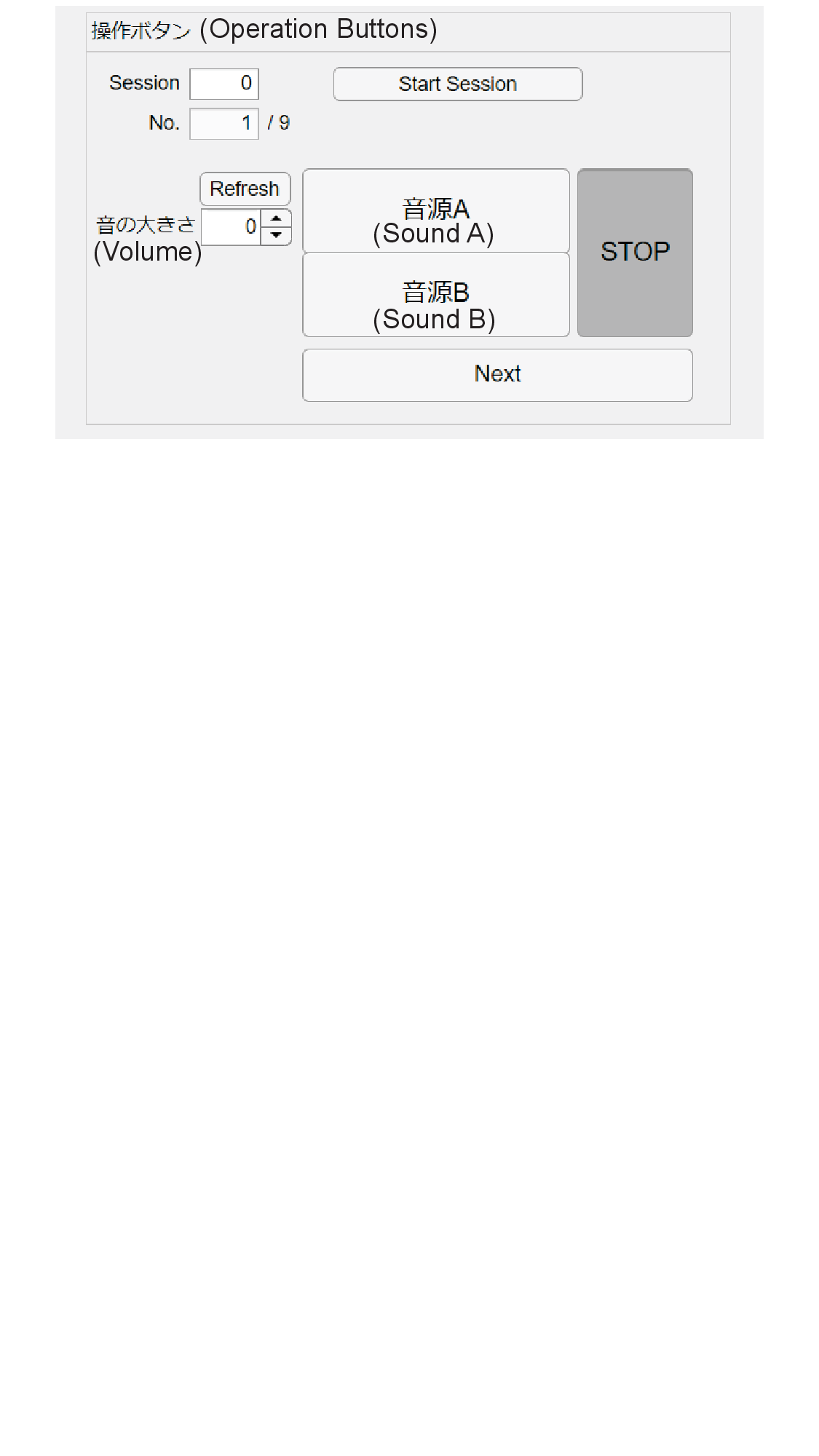}
\vspace{-115mm}
\caption{User interface for the experiment.}
\label{ui}
\end{figure}

\paragraph{User Interface}
\label{sec:ui}
Fig. \ref{ui} presents the user interface used in this experiment, which was created with MATLAB. Pressing the ``Sound A'' button plays the Original signal ($\rm{S_0}$) along with noise ($\rm{N_0}$), and pressing the ``Sound B'' button plays the Comparison signal (variations of $\rm{S_{\pi}}$) along with noise ($\rm{N_0}$)\footnote{Note that in some experimental conditions, both Sounds A and B may play the Original signal. Details are described later in Section \ref{sec:cond}.}. The noise played with Sounds A and B is the same.

\begin{figure}[btp]
\centering
\includegraphics[width=\linewidth]{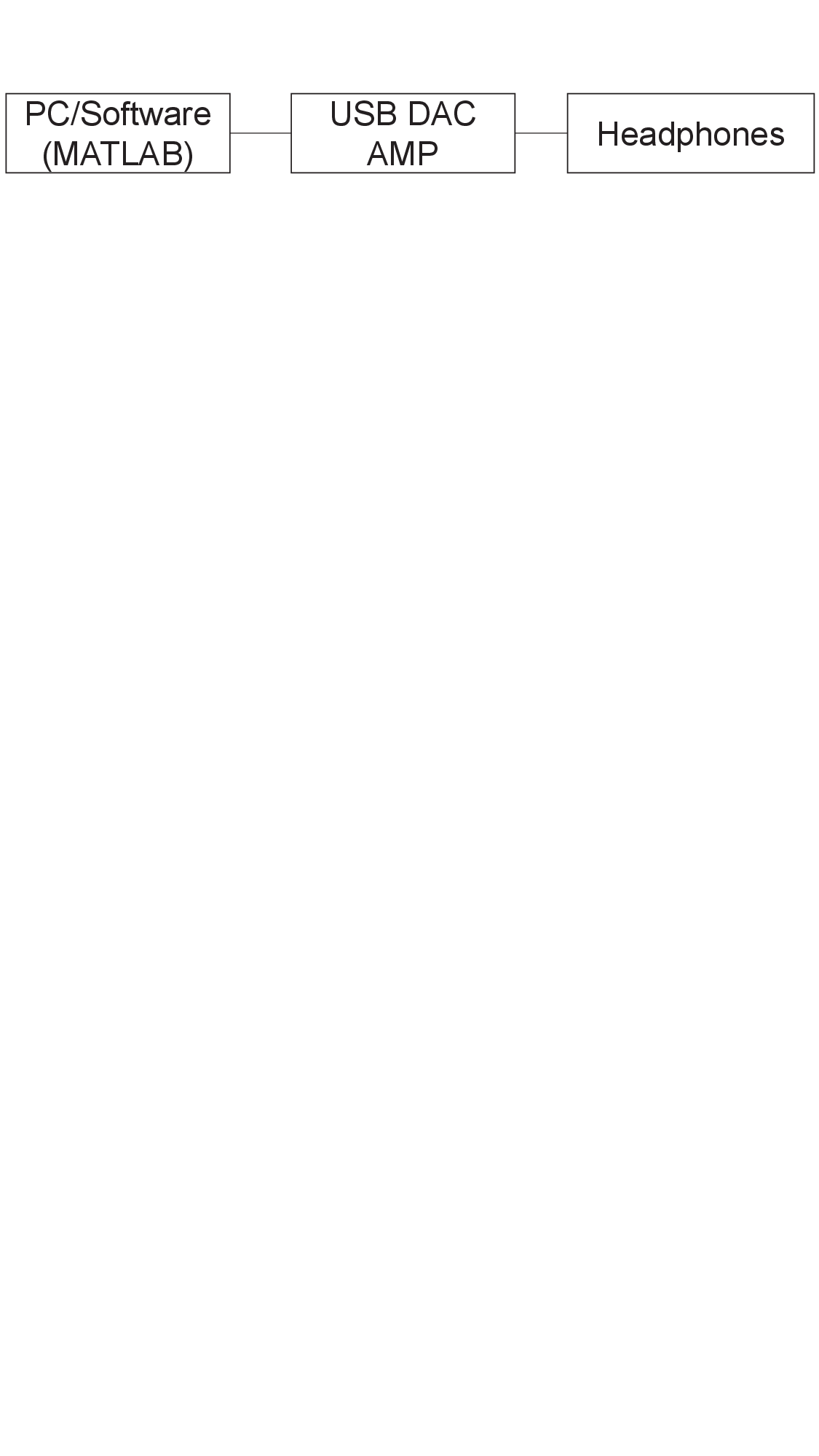}
\vspace{-143mm}
\caption{Layout of equipment.}
\label{equipment}
\end{figure}

\begin{figure*}[btp]
\centering
\includegraphics[width=\linewidth]{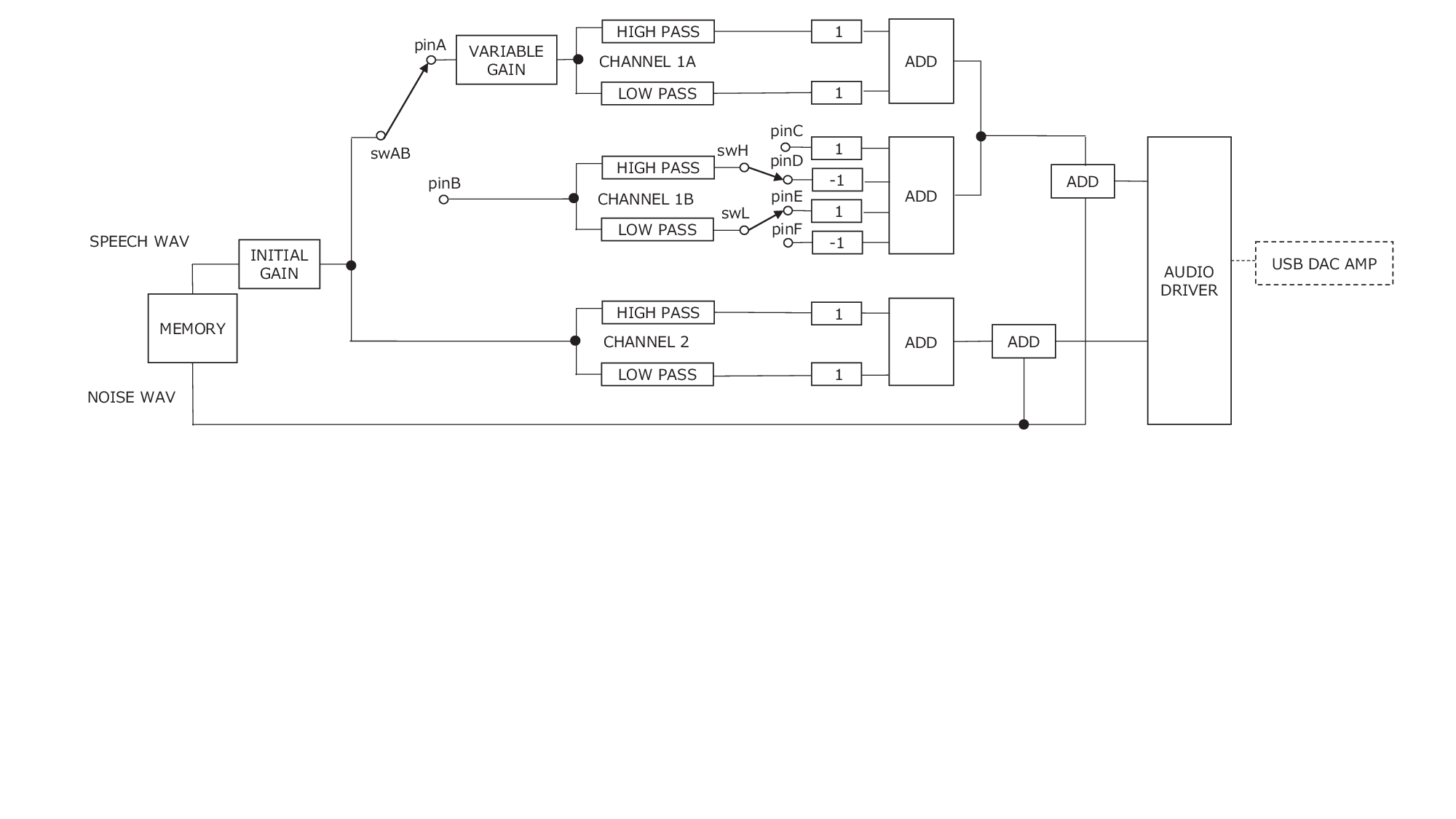}
\vspace{-50mm}
\caption{Block diagram of signal processing in PC.}
\label{sig-pro}
\end{figure*}

Participants compare the two sounds by alternately pressing the ``Sound A'' and ``Sound B'' buttons. Then, they use the spin button to adjust the volume of the speech in Sound A in 1-dB increments until the perceived level of the speech in Sounds A and B feels the same. To prevent bias in participants' responses, the user interface displayed volume as unlabeled numeric values, and participants were not informed that each increment represented 1 dB. Note that only the volume of the speech is adjusted, and the volume of the noise remains unchanged. Once the comparison and volume adjustment are completed, the participant presses the ``Next'' button. This action records the adjusted volume and loads the next stimuli (i.e., Sounds A and B). Simultaneously, the number next to ``No.'' automatically increments.

The stimuli were presented to participants in a random order with special care to avoid bias in the experimental conditions assigned to each participant. Moreover, presenting the stimuli in a random order was a measure to ensure that the fatigue of the participants over time does not affect the results for specific experimental conditions. Further details are provided in Appendix \ref{sec:order}.

\paragraph{Instruction to Participants}
\label{sec:inst}
Participants were instructed to adjust the volume of the speech in Sound A so that it matched the perceived level of the speech in Sound B while alternating playback between the two sounds (i.e., Sounds A and B). The participants were told they could listen to the two sounds repeatedly and switch between them even in the middle of a sentence or word. If participants felt that the level of the speech in Sounds A and B was originally the same, they were instructed not to adjust the volume. They were also asked to ignore any differences in sound qualities other than the level of speech and to focus solely on comparing levels. If they wanted to stop playing the speech and noise, they were told to press the ``STOP'' button. Once the volume adjustment was complete, they were instructed to press the ``Next'' button to proceed to the next stimulus.

In each session, nine comparisons were conducted, and four sessions were carried out per participant. In other words, each participant completed 36 comparisons ($9\times4=36$). A maximum break of about 5 minutes was provided between sessions. Additionally, participants could take breaks at times other than the scheduled rest periods upon request. The total experiment time per participant, including breaks, was less than one hour. Although the one-hour experiment may not have been easy for the participants, it was conducted in accordance with procedures approved by the ethics review board, taking into account the risks to the participants. Therefore, it is believed that the experiment was unlikely to compromise the physical or mental health of the participants. Note that the first of the four sessions served as a practice session and is not included in the results shown in Section \ref{sec:res}.

\subsection{Apparatus and Experimental Environment}
\label{sec:app}
\subsubsection{Devices Used for Sound Presentation}
\label{devices}
Fig. \ref{equipment} shows the layout of the devices used for the experiment. The user interface described in \ref{sec:ui} was run on a Windows PC. The participants used Sony headphones MDR-1AM2\footnote{\url{https://electronics.sony.com/audio/headphones/all-headphones/p/mdr1am2-b}} to listen to the sounds. A Sony WALKMAN NW-A55\footnote{\url{https://www.sony.com/electronics/support/digital-music-players-nw-nwz-a-series/nw-a55}} was connected between the PC and the headphones, and the WALKMAN's USB DAC (Digital to Analog Converter) function was utilized to set the volume displayed on the WALKMAN to 52. The connection between the headphones and the WALKMAN was made using the original cable that comes with the headphones. The signal was output from PC to USB DAC AMP through the Audio Driver (ASIO). Under this device configuration, the A-weighted equivalent continuous sound pressure level (LAeq) remained below 65 dBA, even when the speech volume was adjusted using the user interface. This LAeq is well below levels known to pose a risk to human hearing\cite{who:2019, ISO:1999}.

\subsubsection{Apparatus for Signal Processing}
\label{sec:sig-pro}
Fig. \ref{sig-pro} shows the signal processing used in this experiment. There are two blocks related to volume adjustment: the ``INITIAL GAIN'' block, which determines the initial gain of the signal presented to the participants, and the ``VARIABLE GAIN'' block, which the participants themselves operate to adjust the gain. The sound presented to the left ear (i.e., CHANNEL 1A and 1B) and the sound presented to the right ear (i.e., CHANNEL 2) are each processed through ``HIGH PASS'' and ``LOW PASS'' filters, respectively, then phase reversal (i.e., multiplying by $- 1$) is applied as needed, and the signals are combined in the ``ADD'' block. Details of this process are described in Appendix \ref{sec:method}.

\subsubsection{Rooms Used in Experiment}
The experiment took place in quiet conference rooms (three locations) within the company to which the authors belong. The noise level in both conference rooms was below 40 dBA, which was deemed appropriate for the experiment.

\subsection{Experimental Conditions}
\label{sec:cond}
There were three types of signal and three types of noise, and eight methods of processing the signals, resulting in a total of 72 experimental conditions ($3\times3\times8=72$). The methods of processing the signals included phase reversal of frequency bands below a specific cutoff frequency ($F_c$), phase reversal of frequency bands above $F_c$, and phase reversal across all frequency bands. We refer to these methods as {\em Low Booster}, {\em High Booster}, and {\em All Booster}, respectively.

For the Low/High Booster methods, three different $F_c$ values (250 Hz, 500 Hz, and 1000 Hz) were used. Additionally, the unprocessed signal (i.e., Original) was utilized for comparison. Thus, there were a total of eight types of processing: three Low Boosters, three High Boosters, and one All Booster, plus the Original. Details of the processing methods are described in Appendix \ref{sec:method}.

In conditions where both sounds (i.e., Sounds A and B) were Original, no volume adjustment was expected as there should be no difference in audibility. This condition is referred to as ``Dummy'' hereafter. By examining the results of the Dummy condition, we confirmed whether each participant correctly understood the purpose of the experiment and was able to concentrate on it.

In this experiment, 6 trials were conducted for each of the 72 experimental conditions, resulting in a total of 432 trials, which were divided among the 16 participants.

\begin{table}[btp]
\caption{Results of ``Dummy'' conditions. ``N/A'' of Dummy 4 indicates that the participant performed the Dummy test only three times. Note that P9 was excluded from the aggregation of experimental results due to the large volume adjustment.}
\centering
\begin{tabular}{crrrrr}\bhline{1.5pt}
\multirow{2}{*}{Participant ID} & \multicolumn{4}{c}{BHLD [dB]} \\
& Dummy 1 & Dummy 2 & Dummy 3 & Dummy 4 \\ \hline
P1 & 0 & 0 & 4 & N/A \\
P2 & 0 & 0 & 1 & 0 \\
P3 & 2 & $-1$ & $-1$ & N/A \\
P4 & $-2$ & 0 & 0 & N/A \\
P5 & 1 & 1 & 0 & 2 \\
P6 & $-1$ & 1 & 1 & N/A \\
P7 & 0 & 0 & 0 & 1 \\
P8 & 1 & 0 & 0 & N/A \\
\rowcolor[gray]{0.9}%
P9 & 2 & 5 & 1 & N/A \\
P10 & $-1$ & $-2$ & 2 & 2 \\
P11 & 0 & 0 & 0 & N/A \\
P12 & 1 & 1 & 1 & N/A \\
P13 & 0 & 0 & 1 & 0 \\
P14 & $-1$ & 0 & $-1$ & N/A \\
P15 & 0 & 0 & 0 & 0 \\
P16 & 2 & 2 & 0 & N/A \\
P17 & 1 & 0 & 2 & N/A \\
\bhline{1.5pt}
\end{tabular}
\label{resultdummy}
\end{table}

\section{Results}
\label{sec:res}
In this paper, BHLD  refers to the amount adjusted by the participants using the ``volume'' button, or in other words, the value obtained by subtracting the pre-adjustment volume from the post-adjustment volume. This value indicates the degree of improvement in audibility, that is, how much easier it became to hear the speech with each method.

The results for the Dummy condition (where sound source A and B were identical) are shown in Table \ref{resultdummy}. The values shown represent the BHLD obtained from three to four Dummy trials conducted by each participant\footnote{Each session included one or two dummy trials, and since each participant conducted three sessions, the total was three to four trials. As mentioned in Section \ref{sec:inst}, participants actually conducted four sessions each, but the first session was for practice and is not included in the results tally.}. Note that P9 was excluded from the aggregation of experimental results because the volume adjustment deviated from zero (greater than 4 dB). According to these results, each participant, except for P9, understood the purpose of the experiment and did not make substantial volume adjustments in conditions where there was no difference in audibility. Although participant P1 made a 4-dB adjustment once, the other two trials were maintained at 0 dB, so it was deemed not to be an issue.

\begin{figure}[btp]
\centering
\includegraphics[width=\linewidth]{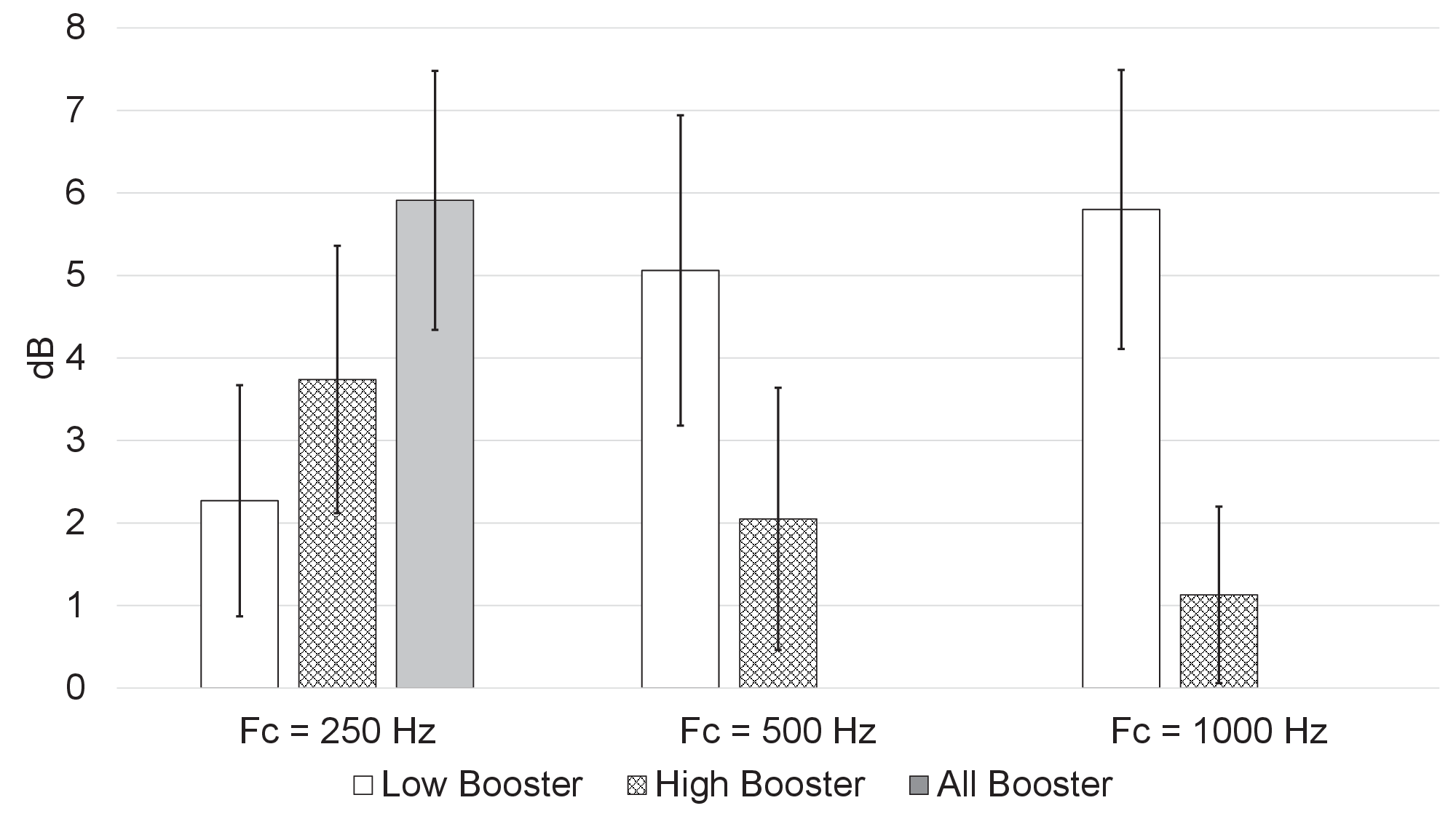}
\vspace{-5mm}
\caption{Experimental results averaged over all signals and all noises. $F_c$ for ``All Booster'' was 250 Hz only. Error bars represent standard deviation.}
\label{resultall}
\end{figure}

\begin{figure}[btp]
\centering
\includegraphics[width=\linewidth]{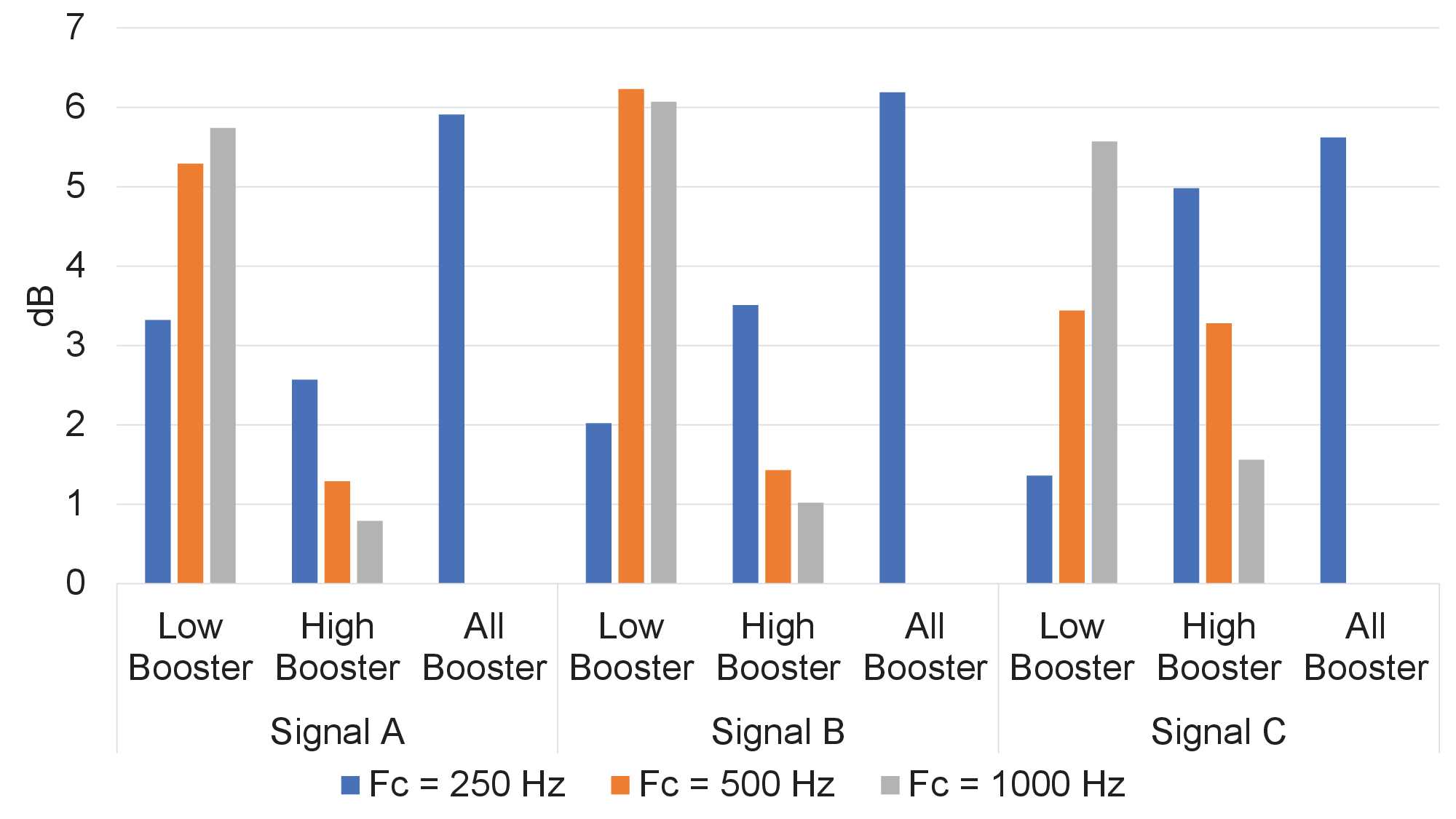}
\vspace{-5mm}
\caption{Experimental results by signal. $F_c$ for ``All Booster'' was 250 Hz only. For visibility, error bars are omitted.}
\label{resultsignal}
\end{figure}

\begin{figure}[btp]
\centering
\includegraphics[width=\linewidth]{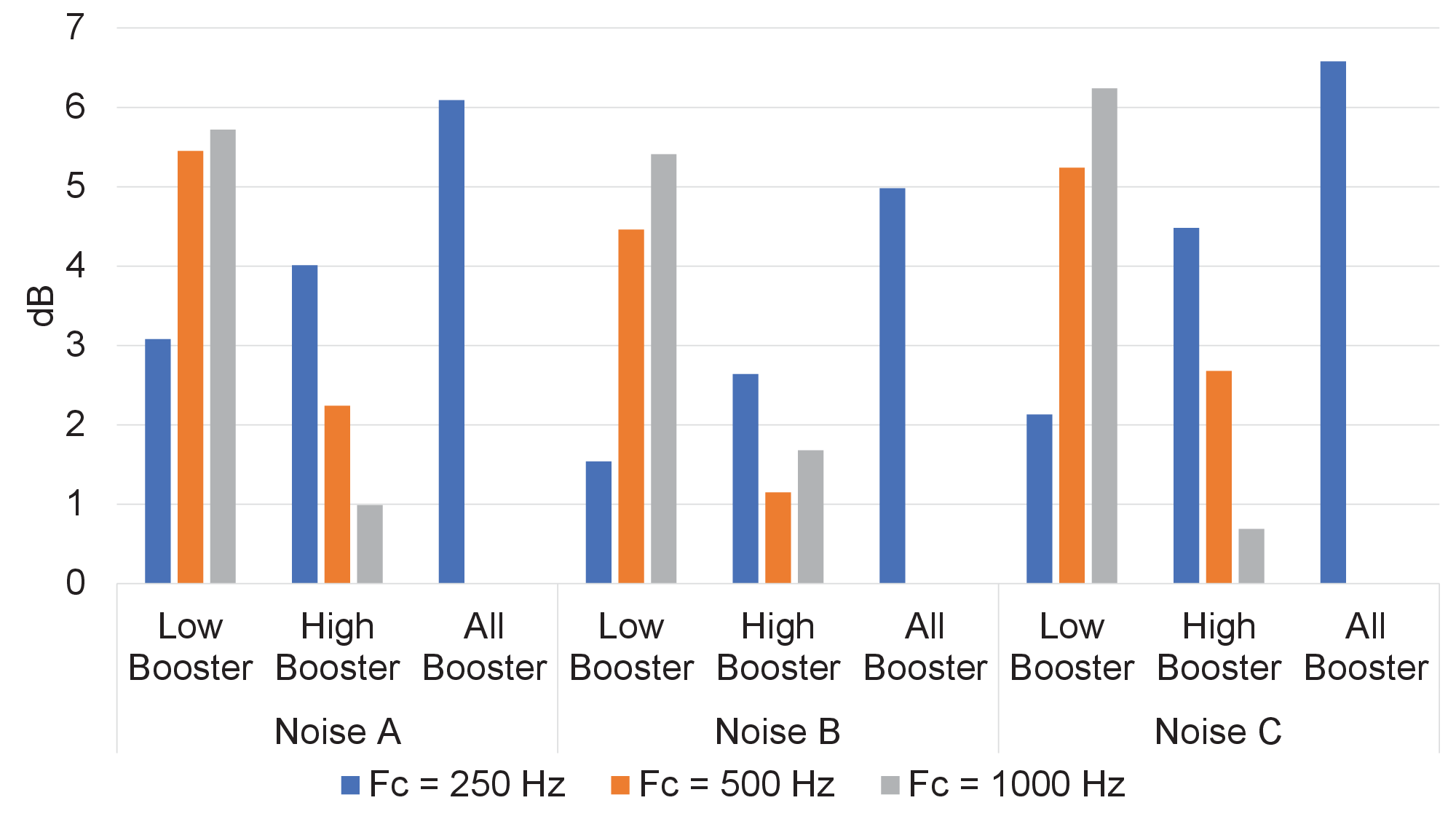}
\vspace{-5mm}
\caption{Experimental results by noise. $F_c$ for ``All Booster'' was 250 Hz only. For visibility, error bars are omitted.}
\label{resultnoise}
\end{figure}

Fig. \ref{resultall} shows the aggregated results for all conditions, and Figs. \ref{resultsignal} and \ref{resultnoise} show the results separated by signal and noise, respectively. In all figures, the vertical axis represents the averages of the BHLD obtained from each trial. For example, the value for ``Low Booster'' at ``$F_c$ = 250 Hz'' in Fig. \ref{resultall} (2.27 dB) is the average of the BHLD from six trials for the corresponding experimental conditions (i.e., the average of 3 signals $\times$ 3 noises $\times$ 6 trials).

In the aggregated results for all conditions (Fig. \ref{resultall}), there was no statistically significant difference between the BHLD for All Booster and Low Booster at $F_c$ = 1000 Hz ($p = 0.75$, Welch's t-test). Both BHLD values were about 6 dB. Specifically, the sample means for All Booster and Low Booster at $F_c$ = 1000 Hz were 5.91 dB and 5.80 dB, respectively. The 95\% confidence intervals (CIs)  based on t-distribution (df = 53) were 5.42 dB to 6.37 dB and 5.27 dB to 6.30 dB for All Booster and Low Booster at $F_c$ = 1000 Hz, respectively, indicating that it is highly likely that the true population means fall within these ranges. For Low Booster, the BHLD increased as $F_c$ increased. On the other hand, for High Booster, the BHLD decreased as $F_c$ increased.

In the results separated by signal (Fig. \ref{resultsignal}), differences were observed depending on the signal. Specifically, while there was no statistically significant difference between the BHLD of $F_c$ = 500 Hz and $F_c$ = 1000 Hz for Low Booster in Signals A and B ($p = 0.49$ and $p = 0.82$, respectively, Welch's t-test), a significant improvement at $F_c$ = 1000 Hz was observed for Signal C ($p < 0.001$, one-sided Welch's t-test), with the BHLD of 3.44 dB and 5.57 dB for $F_c$ = 500 Hz and $F_c$ = 1000 Hz, respectively.

In the results separated by noise (Fig. \ref{resultnoise}), differences were observed depending on the noise. The BHLD reached a maximum of about 6 dB for Noise A and C, whereas for Noise B, the maximum was approximately 5 dB.

For all signals and noises, a BHLD of more than 5 dB was obtained. For All Booster of Signals A, B, and C, the sample means were 5.91 dB, 6.19 dB, and 5.62 dB, respectively, and the 95\% CIs based on t-distribution (df = 17) were 4.96 dB to 6.76 dB, 5.37 dB to 6.93 dB, and 4.62 dB to 6.52 dB, respectively. For All Booster of Noises A, B, and C, the sample means were 6.09 dB, 4.98 dB, and 6.58 dB, respectively, and the 95\% CIs based on t-distribution (df = 17) were 5.24 dB to 6.86 dB, 5.74 dB to 7.35 dB, and 4.11 dB to 5.77 dB, respectively. As for Noise B, Low Booster at $F_c$ = 1000 Hz showed the largest BHLD (5.41 dB), and the 95\% CI based on t-distribution (df = 17) was 4.33 dB to 6.37 dB.

\section{Discussion}
\label{sec:disc}
\subsection{Overall}
The aggregated BHLD for all conditions (Fig. \ref{resultall}) showed a similar trend to the BMLD reported in the previous study \cite{levitt:1967} that focused on English speech read by one male, in the following aspects:

\begin{enumerate}
\item The effect of Low Booster\footnote{Corresponding to $\rm{S\pi0/}fd\rm{N0}$ in the previous study \cite{levitt:1967}.} was greater with a {\bf higher} $F_c$ ($p<0.001$, comparing between $F_c$ = 250 Hz and $F_c$ = 500 Hz, one-sided Welch's t-test). However, there was no significant difference in effect between $F_c$ = 500 Hz and $F_c$ = 1000 Hz ($p = 0.058$, Welch's t-test).
\item The effect of High Booster\footnote{Corresponding to $\rm{S0\pi/}fd\rm{N0}$ in the previous study \cite{levitt:1967}.} was greater with a {\bf lower} $F_c$. In particular, the effect significantly decreased when $F_c$ was raised from 250 Hz to 500 Hz ($p < 0.001$, one-sided Welch's t-test).
\item There was no significant difference in effect between All Booster\footnote{Corresponding to S0N0 in the previous study \cite{levitt:1967}.} and Low Booster at $F_c$ = 1000 Hz\footnote{Corresponding to $\rm{S\pi0/1000N0}$ in the previous study \cite{levitt:1967}.} ($p = 0.75$, Welch's t-test).
\end{enumerate}

\noindent
Items 1) and 2) above demonstrate that both Low Booster and High Booster achieved greater effects when a wider frequency band was phase-inverted. However, narrowing the phase-inverted frequency band did not necessarily lead to a decrease in effect. Specifically, there was not much difference in the effect when the inverted frequency band for Low Booster was narrowed from below 1000 Hz to below 500 Hz. Additionally, item 3) indicates that phase-inverting a limited frequency band can achieve the same level of audibility as inverting the entire frequency band in one ear's signal.

While similar trends to the previous study \cite{levitt:1967} were observed in these aspects, the magnitude of the effects differed. Specifically, the BMLD obtained in the previous research \cite{levitt:1967} was up to 14 dB, whereas the BHLD obtained in this study was up to approximately 6 dB. This difference may be due to the difference in evaluation methods (refer to Section \ref{sec:purpose} for the description). We leave the quantitative analysis of the differences in evaluation methods to future research.

In terms of the effect on speaker variation, a consistent effect (i.e., BHLD of at least 5.62 dB) was obtained for all speakers. This means that under the conditions of this experiment, there were no speakers for whom it was particularly difficult to achieve an effect of hearing improvement.

Regarding the effect on noise variation, a relatively large difference in BHLD was observed among noises. However, since a BHLD of more than 5 dB was obtained for all noises, we conclude that under the conditions of this experiment, there were no noises for which it was particularly difficult to achieve an effect of hearing improvement.

\subsection{Detailed Analysis by Signal and Noise}
\label{detail}
\begin{figure*}[btp]
\centering
\includegraphics[width=\linewidth]{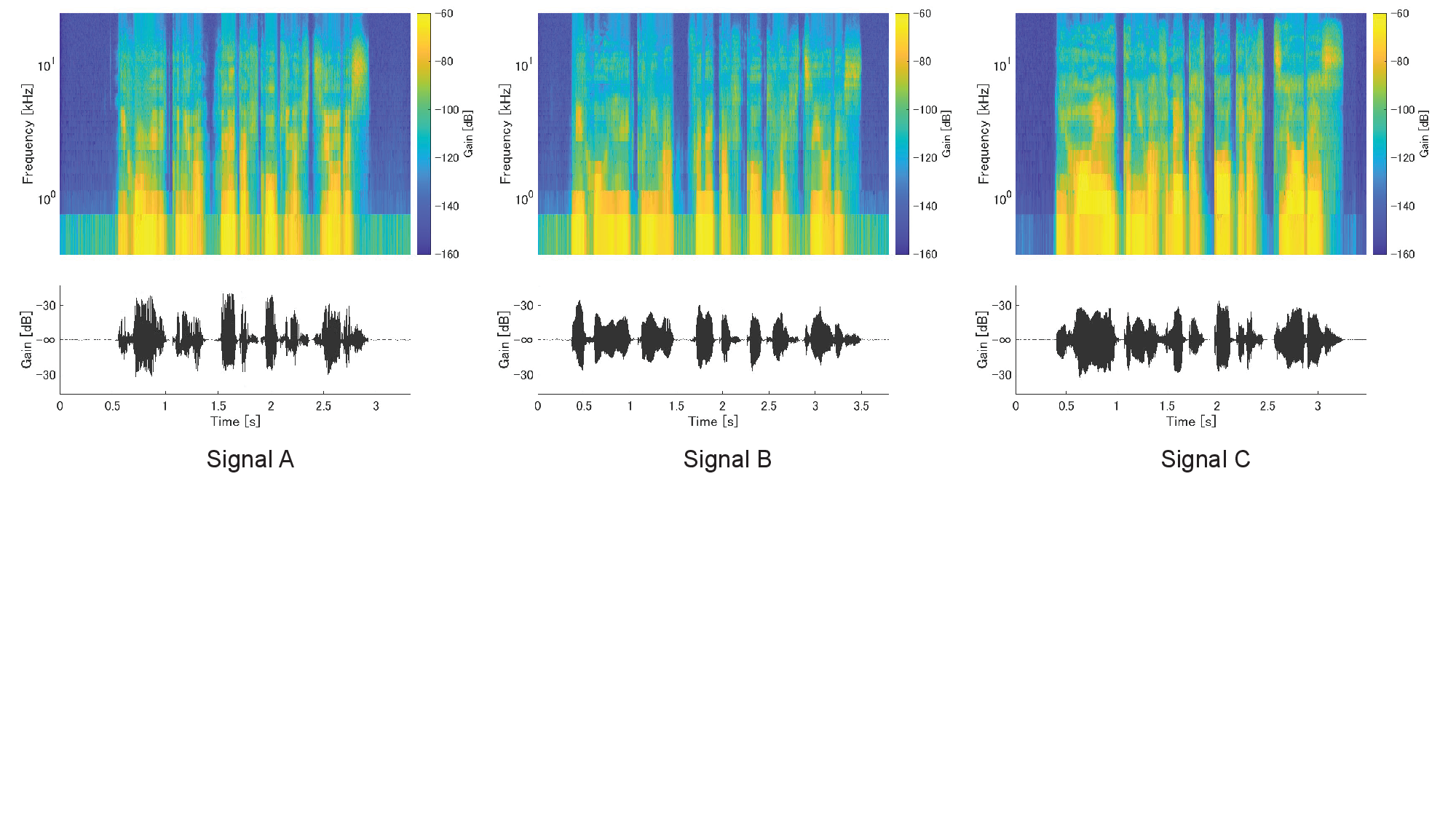}
\vspace{-45mm}
\caption{Spectrograms and waveforms of Signals A, B, and C. The upper part corresponds to the spectrogram and the lower part to the waveform.}
\label{spec-signal}
\end{figure*}

\begin{figure*}[btp]
\centering
\includegraphics[width=\linewidth]{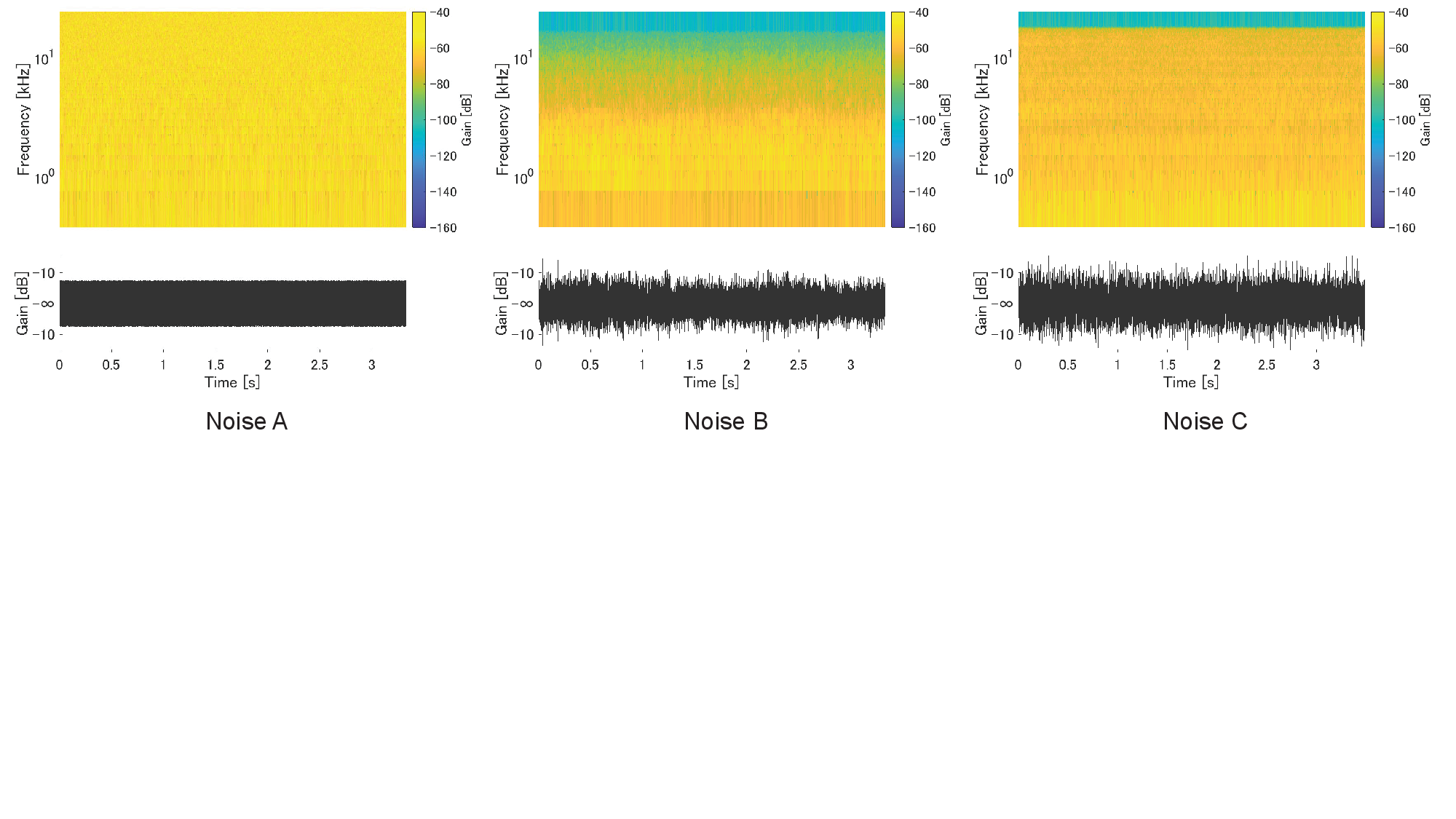}
\vspace{-50mm}
\caption{Spectrograms and waveforms of Noises A, B, and C. The upper part corresponds to the spectrogram and the lower part to the waveform.}
\label{spec-noise}
\end{figure*}

Following the above experiment, we analyzed  the BHLD aggregated by Signal or Noise (Figs. \ref{resultsignal}, \ref{resultnoise}) considering the spectrograms (Figs. \ref{spec-signal}, \ref{spec-noise}). Note that the spectrograms and waveforms in Fig. \ref{spec-signal} are from the speech with the initial gains when combined with Noise A (i.e., rows 1, 4, and 7 of Table \ref{level} in Appendix \ref{sec:initgain}).

Fig. \ref{resultsignal} shows that the BHLD values of Signal C with High Booster are larger than those of Signals A and B. Specifically, in the case of $F_c$ = 250 Hz, the BHLD of Signal C is significantly higher than that of Signals A and B ($p<0.001$ and $p=0.003$ when compared to Signals A and B, respectively, one-sided Welch's t-test).

Comparing the spectrograms of Signals A, B, and C in Fig. \ref{spec-signal}, we observe that Signal C had stronger high-frequency components than Signals A and B. This could explain why the BHLD for High Booster in Signal C was larger compared to Signals A and B.

The reason the BHLD for Noises A and C exhibited similar trends in Fig. \ref{resultnoise} is presumably due to both Noises A and C having a uniform frequency distribution. In contrast, the BHLD for Noise B showed different tendencies, as follows.

\begin{enumerate}
\item For Low Booster, the BHLD at $F_c$ = 250 Hz is relatively smaller compared to other Noises.
\item For High Booster, the BHLD does not change substantially with the changes in $F_c$.
\end{enumerate}

\begin{figure*}[btp]
\centering
\includegraphics[width=\linewidth]{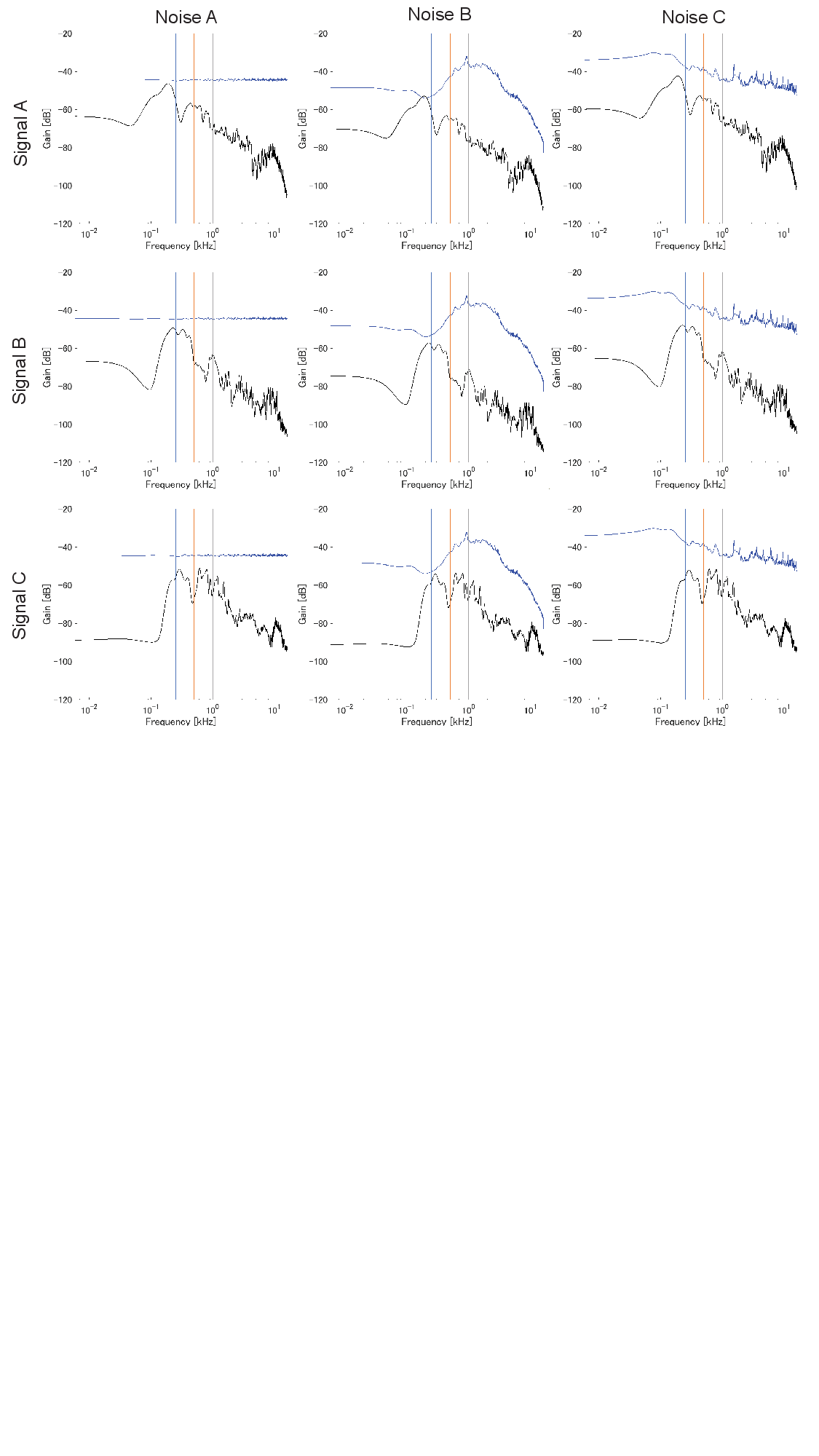}
\vspace{-160mm}
\caption{Relationships between Signal and Noise gains. Blue and black plots correspond to the gains of noises and signals, respectively. Vertical lines in blue, orange, and gray correspond to 250, 500, and 1000 Hz, respectively.}
\label{gain}
\end{figure*}

We analyzed the above trends based on the relationship between the gains of the Signal and Noise shown in Fig. \ref{gain}. The gain of Noise B gradually increased from around 250 Hz, peaked near 1000 Hz, and then attenuated gradually at frequencies higher than 1000 Hz. The gains of Signals A, B, and C each had peaks around 200 Hz, 200–300 Hz, and 300 Hz as well as 600–800 Hz, respectively, and then attenuated gradually beyond those frequencies. The main frequency components of Noise B were between 800–2000 Hz, which are on a higher frequency compared to the peaks of Signals A, B, and C.

Regarding trend 1), it seems that the frequency band processed with Low Booster (i.e., below 250 Hz) became more audible for all Noises. However, in the case of Noise B, the masking in the frequency band above 250 Hz was stronger than in Noises A and C, resulting in a difficulty in hearing overall. This is presumably why the BHLD of Noise B was smaller compared to that of Noises A and C.

Regarding trend 2), although the frequency bands processed with High Booster (i.e., over 250 Hz, 500 Hz, and 1000 Hz) became more audible than Original, the case of Noise B became more difficult to hear compared to the cases of Noises A and C, resulting in a smaller BHLD. This is because the gain attenuation of the Signals and the gain increase of Noise B occurred simultaneously in these frequency bands.

These analyses indicate that the BHLD varies depending on the frequency distributions of both the target signal and the noise. This means there is a possibility of achieving a higher BHLD by conducting parameter exploration (i.e., investigating the types of Boosters and the values of $F_c$) according to the frequency distributions of both the target sound and noise.

\subsection{Values and Concerns for Practical Applications}
Some of the participants reported feeling a change in localization under anti-phasic presentation. This is consistent with the theory that the signal becomes easier to detect in noise due to a subtle shift in the sound image caused by phase reversal \cite{mcfadden:1971}, and with previous findings that the target sound becomes more audible as a result of the brainstem's SOC localizing the target sound at a different position from the noise due to the interaural phase difference of the target sound \cite{jiang:1997}. On the other hand, another study reported that a target sound 180\textdegree out-of-phase interaurally does not make the perceived localization of the target sound and noise clearly separate; rather, it makes the sound images perceived as diffused inside the head \cite{carhart:1969}. The diffusion of sound images is a phenomenon that does not occur in everyday situations and may therefore cause a particularly strong sense of discomfort.

Considering practical use, it is necessary to reduce the discomfort in auditory perception. Based on the discussion in \ref{detail}, we expect that by limiting the frequency band of phase reversal in one ear, a more effective BHLD can be achieved while avoiding the discomfort in sound localization caused by inverting the entire frequency range.

We consider the experimental results of this paper to be worthwhile from the perspective of solving social issues. Safe listening levels depend on the intensity and duration of exposure \cite{who:2015}: for example, if one listens to sound at 89 dB (assuming listening through earphones in noisy conditions), according to the WHO-ITU standard \cite{who:2019}, the permissible listening time for auditory safety over the course of one week is five hours. Using a Booster can make the volume feel about 6 dB louder, so for example, when the actual sound pressure is 83 dB, the same level of audibility as 89 dB can be achieved. According to literature \cite{who:2019}, a reduction in volume from 89 dB to 83 dB corresponds to an increase of 15 hours in the permissible listening time for adults over one week. Therefore, the 6-dB hearing improvement can be said to have a significant social impact.

\section{Conclusion}
In this study, we investigated the effects of binaural unmasking through various anti-phasic conditions, focusing on variations in speakers and noises, with the aim of applying the phenomenon to engineering applications. The experimental results targeting the Japanese language confirmed that anti-phasic presentation can make a speech, which was barely audible in a noisy environment, be perceived as up to about 6 dB louder. An improvement of 6 dB in perceived level implies that, after phase reversal in one ear, the sound pressure can be reduced by 6 dB while maintaining the same perceived loudness as before the phase reversal. Reducing the sound pressure level by 6 dB increases the permissible safe listening time by about 15 hours per week under a given sound pressure condition. Although the degree of effect obtained varied depending on  the target signals and the noises, a consistent effect (BHLD of at least 5 dB) was achieved for all target signals and noises addressed in this study.

\appendices

\section{List of Online Resources}
Table \ref{resourcelist} lists the online resources that were utilized in this paper.

\begin{table*}[btp]
\caption{List of online resources utilized in this paper.}
\small
\centering
\begin{tabular}{p{4.3cm}p{2cm}p{3.8cm}p{5cm}}\bhline{1.5pt}
Resource & Section of first appearance & License & URL \\ \hline
ITA Corpus & \ref{sec:signal} & Unlicense license & \url{https://github.com/mmorise/ita-corpus} \\
SpeedSpeech-JA-2022 & \ref{sec:signal} & CC BY-NC 4.0 \par\noindent (\url{https://creativecommons.org/licenses/by-nc/4.0/deed.en}) & \url{https://ast-astrec.nict.go.jp/release/speedspeech_ja_2022/} \\
Amitaro's Voice Material Studio & \ref{sec:signal} & Always provide credit when using it for free. & \url{https://amitaro.net/} \\
TK'S FREE SOUND FX & \ref{sec:noise} & Allowed to use the sounds free of charge and royalty free. & \url{https://taira-komori.jpn.org/freesounden.html} \\
OtoLogic & \ref{sec:noise} & CC BY 4.0 \par\noindent (\url{https://creativecommons.org/licenses/by/4.0/deed.en}) & \url{https://otologic.jp/} \\
\bhline{1.5pt}
\end{tabular}
\label{resourcelist}
\end{table*}

\section{Preparation of Noise A}
\label{sec:noisea}
Uniformly distributed random numbers were generated and recorded at a sampling frequency of 48 kHz. The maximum amplitude was adjusted to $-6$ dB.

\section{Preparation of Noise B}
\label{sec:noiseb}
The processing of the sound source was done using SOUND FORGE Pro 14.0\footnote{Product information can be found at the following URL, which is for the latest version (SOUND FORGE Pro 18): \url{https://www.magix.com/gb/music/sound-forge/sound-forge-pro/}}. First, we downloaded the ``Cheer Crowd01'' folder from the OtoLogic website\footnote{\url{https://otologic.jp/free/se/applause-cheer01.html}} and selected \texttt{Cheer-Crowd01-1(Low-Long).mp3}. The reason for choosing this sound source was that it had frequency components that could easily mask the target Signals and was voiced continuously.

After converting this sound source to a monaural signal by averaging between the left and right channels, a segment where the sound pressure is as constant as possible was extracted. To match the sampling frequency of the Signals, the sound source was resampled to 48 kHz and saved in WAV format.

\section{Preparation of Noise C}
\label{sec:noisec}
To simulate noises commonly heard in daily life, we selected urban environmental sounds based on the following criteria: containing low-frequency components to sufficiently mask speech, having a broad frequency distribution, maintaining a consistent sound pressure over a certain period of time, and involving multiple types of sounds occurring simultaneously.

We used five types of sounds from TK'S FREE SOUND FX\footnote{\url{https://taira-komori.jpn.org/enviroment02.html}}: \texttt{building\_home2.mp3}, \texttt{power\_drill1.mp3}, \texttt{drill\_concrete.mp3}, \texttt{construction\_site.mp3}, and \texttt{elct\_rotary\_cutter2.mp3}. These were converted to monaural  by averaging the left and right channels, and segments where the sound pressure was as constant as possible were extracted. These segments are referred to as Set 1.

Next, to ensure that each sound source in Set 1 was added with equal energy, we adjusted the RMS levels of all the sound sources to match the lowest RMS in Set 1. They were then summed and normalized so that the peak value became $-1$ dB. To match the sampling frequency of the Signals, the sound sources were resampled to 48 kHz and saved in WAV format.

\section{Process for Aligning Volume of Noises}
\label{sec:align-noise}

We equalized the signal levels among Noises A, B, and C by performing adjustments utilizing the function preset in SOUND FORGE Pro 14.0 that normalizes the average RMS level. In the experiment, the noises were played for the same duration as the signals (approximately 3 seconds). Specifically, the first approximately 3 seconds of each sound source described in Appendix \ref{sec:noisea} to \ref{sec:noisec} were used. We also confirmed that the segments could be played repeatedly without sounding discontinuous. Dynamic compression was applied when clipping occurred. As shown in Table \ref{align-noise}, after applying this operation, the RMS levels of Noises A, B, and C were $-8.0$ dB, $-7.6$ dB, and $-8.0$ dB, respectively. The spectrograms of Noises A, B, and C after volume adjustment are shown in Fig. \ref{spec-noise}.

\begin{table}[btp]
\caption{Peak and RMS of noises before and after the alignment process.}
\centering
\begin{tabular}{ccrr}\bhline{1.5pt}
 & & Before & After \\
 & & \multicolumn{2}{c}{[dB]} \\ \hline
Noise A & Peak & $-6.0$ & $-6.4$ \\
 & RMS & $-7.5$ & $-8.0$ \\ \hline
Noise B & Peak & $-9.7$ & $-0.7$ \\
 & RMS & $-16.5$ & $-7.6$ \\ \hline
Noise C & Peak & $-2.3$ & $-0.1$ \\
 & RMS & $-10.4$ & $-8.0$ \\
\bhline{1.5pt}
\end{tabular}
\label{align-noise}
\end{table}

\section{Process for Aligning Volume of Speech Signals}
\label{align-speech}
The gains of the speech signals were adjusted for uniformity utilizing the function of normalizing the average RMS level, which is preset in SOUND FORGE Pro 14.0. The average RMS level was set to approximately $-14.0$ dB. Dynamic compression was applied when clipping occurred. As shown in Table \ref{align-signal}, after applying this operation, the RMS levels of Signals A, B, and C were $-14.9$ dB, $-14.5$ dB, and $-14.0$ dB, respectively. 

\section{Determination of Initial Gains of Signals}
\label{sec:initgain}
Considering that listening difficulties in everyday life typically occur when the SNR is negative, conditions with positive SNR were excluded from the experiment. Under positive SNR, the target sound in our stimulus set remained clearly perceptible and was never fully masked by noise.

Furthermore, even when the SNR value was identical across stimulus sets, the degree of listening difficulty varied depending on the combination of target sound and noise. For this reason, the initial gain was adjusted based on human perceptual judgment. 

The initial gain settings were determined by the authors, who did not participate in the experiment. Although the authors' prior knowledge of the target phrases might have facilitated speech perception in noise, this does not reduce the listening difficulty for participants without such knowledge. Therefore, this methodological aspect was considered not to affect the experimental conditions.

The initial gains of the Original signals at the start of the experiment are shown in Table \ref{level}. These gains are the average of the levels determined by three authors who listened to nine types of stimuli (3 types of speech $\times$ 3 types of noise) and judged them to be the minimum volume at which the presence of a speech in the noise is perceptible. This volume setting is intended to create an initial state of ``audible but difficult to hear.'' Note that the speech signals that the authors listened to here had undergone volume alignment (i.e., normalization of the average RMS), as described in Appendix \ref{align-speech}. 

The SNRs after setting the initial gains of the Original signals are shown in Table \ref{snr}. In all conditions, the SNRs exhibited negative values.

\begin{table}[btp]
\caption{Peak and RMS of speech signals before and after the alignment process.}
\centering
\begin{tabular}{ccrr}\bhline{1.5pt}
 & & Before & After \\
 & & \multicolumn{2}{c}{[dB]} \\ \hline
Signal A & Peak & $-19.1$ & $-0.1$ \\
 & RMS & $-34.4$ & $-14.9$ \\ \hline
Signal B & Peak & $-14.1$ & $-0.1$ \\
 & RMS & $-29.1$ & $-14.5$ \\ \hline
Signal C & Peak & $-13.6$ & $-3.3$ \\
 & RMS & $-24.3$ & $-14.0$ \\
\bhline{1.5pt}
\end{tabular}
\label{align-signal}
\end{table}

\begin{table}[btp]
\caption{Initial gains of speech signals for each noise. An initial gain represents the difference from the gain after the RMS adjustment shown in Table \ref{align-signal}.}
\centering
\begin{tabular}{ccc}\bhline{1.5pt}
Signal & Noise & Initial gain [dB] \\ \hline
Signal A & Noise A & $-21.2$ \\
Signal A & Noise B & $-27.7$ \\
Signal A & Noise C & $-17.1$ \\ \hline
Signal B & Noise A & $-22.6$ \\
Signal B & Noise B & $-30.5$ \\
Signal B & Noise C & $-21.1$ \\ \hline
Signal C & Noise A & $-19.7$ \\
Signal C & Noise B & $-22.1$ \\
Signal C & Noise C & $-20.0$ \\
\bhline{1.5pt}
\end{tabular}
\label{level}
\end{table}

\begin{table}[btp]
\caption{SNRs of the stimuli after initial gain adjustment.}
\centering
\begin{tabular}{ccc}\bhline{1.5pt}
Signal & Noise & SNR [dB] \\ \hline
Signal A & Noise A & $-25.1$ \\
Signal A & Noise B & $-28.8$ \\
Signal A & Noise C & $-19.1$ \\ \hline
Signal B & Noise A & $-26.1$ \\
Signal B & Noise B & $-31.1$ \\
Signal B & Noise C & $-22.7$ \\ \hline
Signal C & Noise A & $-25.1$ \\
Signal C & Noise B & $-24.6$ \\
Signal C & Noise C & $-23.5$ \\
\bhline{1.5pt}
\end{tabular}
\label{snr}
\end{table}

\section{Order of Presentation of Stimuli}
\label{sec:order}
\begin{figure}[btp]
\centering
\includegraphics[width=\linewidth]{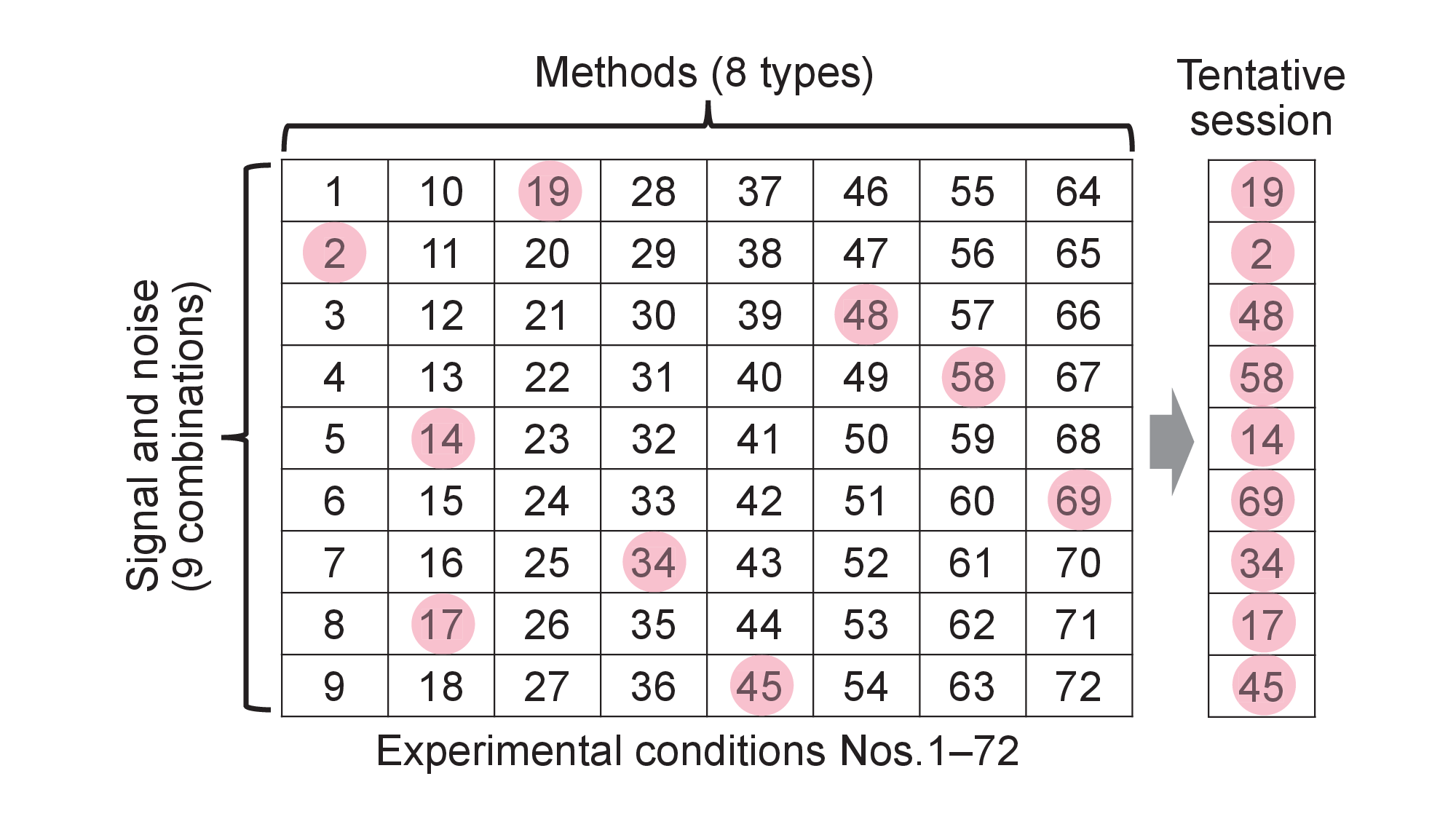}
\vspace{-8mm}
\caption{Example of making a tentative session.}
\label{tmpsession}
\end{figure}

The stimuli corresponding to the experimental conditions outlined in Section \ref{sec:cond} were presented to participants in a random order. However, it was not completely random---the following two steps were implemented to avoid bias in the conditions assigned to each participant:

\begin{enumerate}
\item Stimulus selection\mbox{}\\
Each participant was assigned stimuli for 9 trials $\times$ 3 sessions\footnote{Since one of the four sessions presented to each participant was separately prepared for practice, only three sessions need to be prepared here.}. The sessions created in this step are hereafter referred to as ``tentative sessions.'' Each of the nine combinations of signal and noise (3 signals $\times$ 3 noises) was included once in one tentative session. To avoid bias in methods, eight methods were selected plus one repetition within a tentative session. An example of creating a tentative session is shown in Fig. \ref{tmpsession}. Eight tentative sessions covered all 72 experimental conditions (9 trials $\times$ 8 sessions).
\item Randomization across sessions\mbox{}\\
Out of the eight tentative sessions created in step (1), three were assigned to each participant, and the total of 27 stimuli (9 trials $\times$ 3 sessions) were randomly ordered and presented to the participants. 
\end{enumerate}

\section{Process of Phase Reversal}
\label{sec:method}
\subsection{Overview}
\label{sec:method:overview}
To limit the phase reversal of the signal to certain frequency bands in one ear's channel, the signal was band-split at the boundary defined by $F_c$. To prevent phase misalignment between channels, band-splitting was performed on the signals of both the left and right channels.

The process of phase reversal is explained with reference to Fig. \ref{sig-pro}. Phase reversal was applied to the signal of CHANNEL 1B when Sound B was selected, that is, when swAB was connected to pinB. The switching of the bands for phase reversal was performed by changing the connection terminals of swH and swL. For All Booster, the terminals were changed to [pinD, pinF], for Low Booster, to [pinC, pinF], for High Booster, to [pinD, pinE], and for Original, to [pinC, pinE]. In other words, the coefficients of the output from the Low Pass Filter (LPF) and High Pass Filter (HPF) were switched to [$-1$, $-1$] for All Booster, [$-1$, 1] for Low Booster, [1, $-1$] for High Booster, and [1, 1] for Original. After adding these two bands of signals, the signal was output to USB DAC AMP through the Audio Driver (ASIO). Since no phase reversal was applied to the signal of the right channel (i.e., CHANNEL 2), the signal was split into low- and high-frequency components, but they were added with the coefficients [1, 1]. The right channel was also split to match the time delay caused by filtering the signal of the left channel.

\begin{figure*}[btp]
\centering
\includegraphics[width=\linewidth]{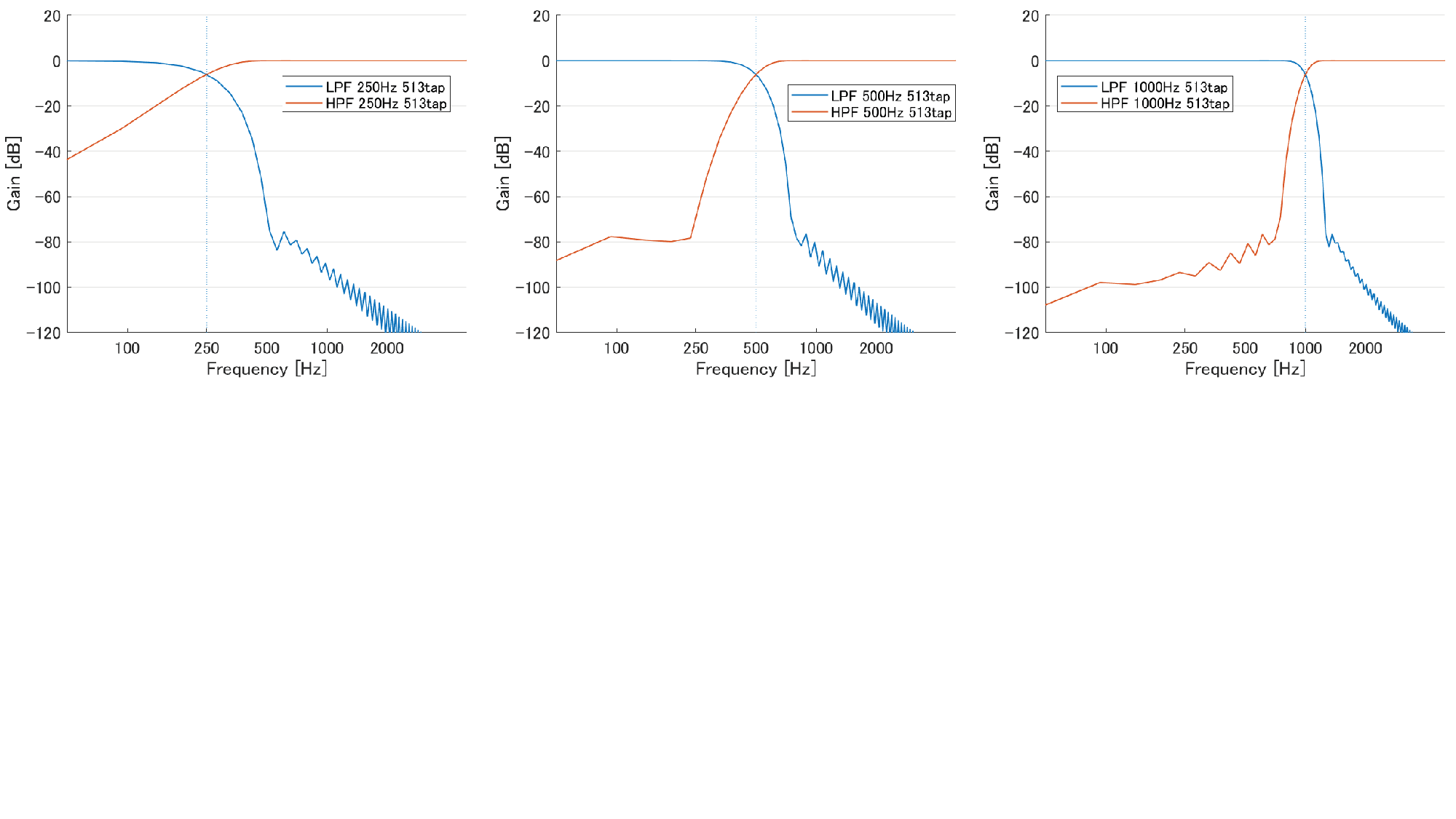}
\vspace{-57mm}
\caption{Filter responses for a tap length of 513. Figures on the left, center, and right correspond to the cases of $F_c$ = 250, 500, and 1000, respectively.}
\label{filter513}
\end{figure*}

\begin{figure*}[btp]
\centering
\includegraphics[width=\linewidth]{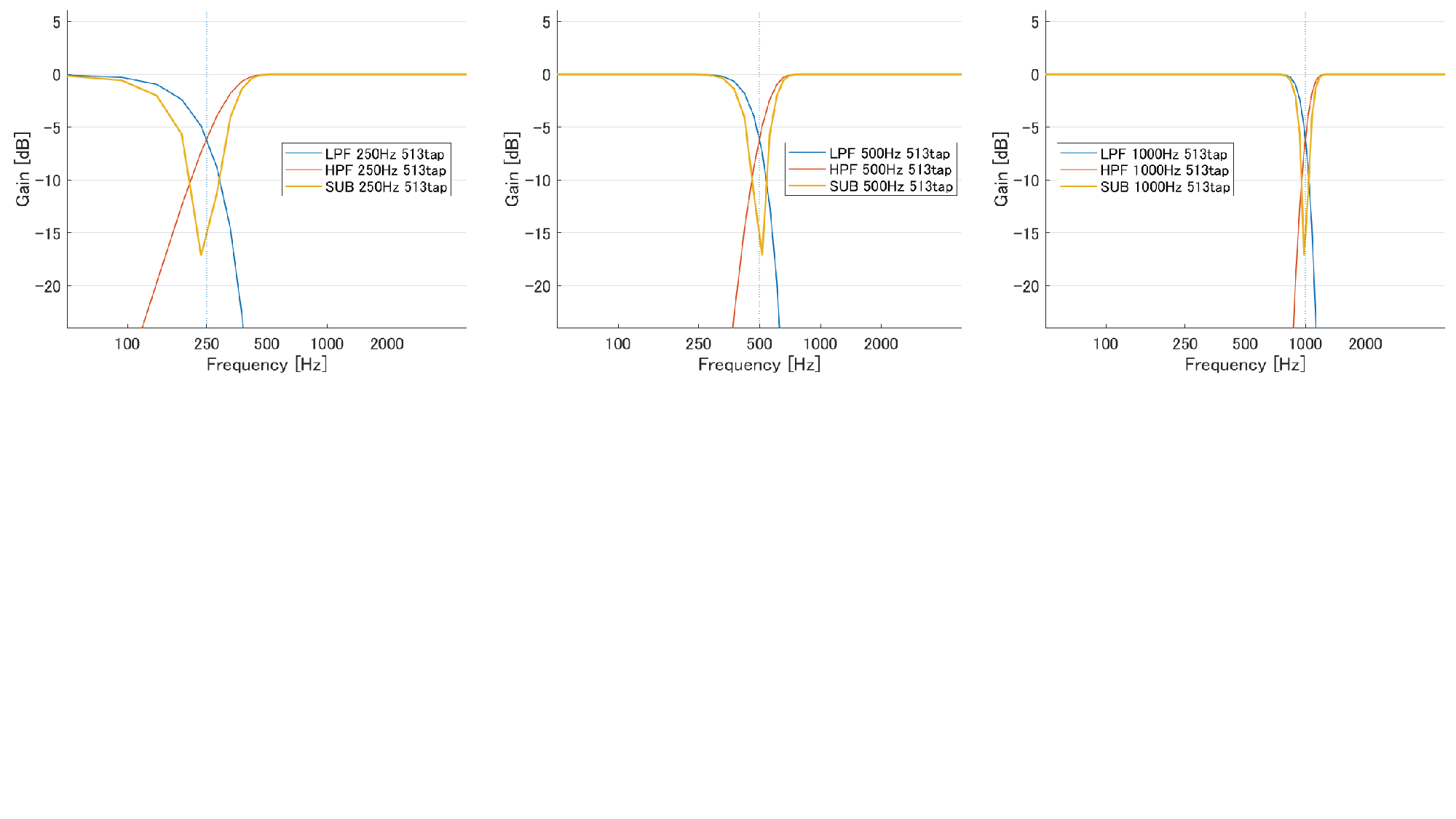}
\vspace{-57mm}
\caption{Volume reduction when either the LPF or HPF is phase-inverted and summed, with a tap length of 513.}
\label{dip513}
\end{figure*}

\subsection{Details of Filter Design}
\label{sec:method:detail}
The LPF and HPF shown in Fig. \ref{sig-pro} were designed as FIR filters using the window function method. The design criteria were as follows:

\begin{itemize}
\item Window type: Blackman
\item Stopband attenuation: About 60 dB
\item Tap length: $2^n+1$
\end{itemize}

\noindent
The tap length can be arbitrary, but we set it to a power of 2 plus 1 in order to narrow down the patterns to be examined for the balance between the filter delay and the frequency response when the outputs of the LPF and HPF were added. The addition of 1 was to make the filter delay an integer. We chose 513 as the tap length because, as shown in Fig. \ref{filter513}, the stopband attenuation of the HPF was about 60 dB even in the case of $F_c$ = 250 Hz. When the input signal was formatted as 16-bit PCM, it was confirmed that adding the outputs of LPF and HPF with tap lengths of 513 without phase reversal resulted in the complete restoration of the original signal. However, when one of the LPF or HPF outputs was phase-inverted (i.e., with coefficients of [1, $-1$] or [$-1$, 1] as in Fig. \ref{sig-pro}) and those outputs were added together, a steep drop in gain occurred near $F_c$ (hereafter referred to as a ``dip''). The dip with a tap length of 513 is shown in Fig. \ref{dip513}.

The dip could potentially affect the BHLD value. A longer tap length (e.g., of 1025) is preferable for having a relatively smaller dip, but 513 is more desirable for having a shorter processing delay. Considering practicality, prioritizing a shorter processing delay is deemed important, and therefore, a tap length of 513 was adopted for this experiment.

\bibliography{202410}
\bibliographystyle{IEEEtran}

\end{document}